\documentclass{revtex4}
\usepackage{graphicx}
\usepackage{fancyhdr}
\usepackage{amsmath}
\usepackage{color}

\begin{document}

\title{SUSY Extended Higgs Sector and SUSY Strong Dynamics}

%

\author{Toshifumi Yamada}
\affiliation{Department of Physics, University of Tokyo, \\
7-3-1 Hongo, Bunkyo-ku, Tokyo 113-0033, JAPAN}

\begin{abstract}
We consider an extended Higgs sector that arises as 
 a low-energy description of a strongly-coupled SUSY gauge theory.
This model predicts the presence of extra superfields that 
 couple with the Higgs superfields through large coupling constants.
Large radiative corrections involving the extra fields can realize 
 strongly first order electroweak phase transition, which is a vital requirement for electroweak baryogenesis scenario.
At the same time, the large radiative corrections lead to significant deviations 
 in the Higgs boson triple coupling and Higgs-to-diphoton branching ratio, which are testable at the ILC.
\end{abstract}

\maketitle

\thispagestyle{fancy}


\section{Introduction}
Although the Higgs-like boson has been discovered,
 we have not yet revealed the structure of the Higgs potential.
In building models of the Higgs sector,
 we adopt electroweak baryogenesis scenario as a guiding principle.
It is known that successful electroweak baryogenesis~\cite{ewbg} 
 relies on sufficient amount of CP violation and strongly first order 
 electroweak phase transition (EWPT),
 both of which are not realized in the standard model (SM) with 126 GeV Higgs boson mass.
The latter condition on EWPT is directly connected to the structure of the Higgs potential
 and may give a clue to the study on the Higgs sector.
We therefore discuss extensions of the SM Higgs sector
 that enable strongly first order EWPT.
In general, models that realize strongly first order EWPT
 through enhanced thermal cubic term necessarily contain large coupling constants in the Higgs sector,
 which blow up at a Landau pole below the Planck scale.
Extended Higgs models with large coupling constants in the Higgs sector have been
 investigated in refs.~\cite{large coupling}.
In such cases, the models
 must be replaced by more fundamental theories above the Landau pole.

We propose an ultraviolet (UV) complete model of an extended Higgs sector
 that incorporates such large coupling constants \cite{ours1, ours2}.
Our model is based on supersymmetric (SUSY) SU(2) gauge theory with six doublets,
 similar to the minimal fat Higgs model \cite{fat higgs}.
This gauge theory becomes strongly-coupled at an infrared (IR) scale, which we call
 `confinement scale',
 and below that scale, the theory is described in terms of mesonic superfields
 with an emergent effective superpotential which contains large coupling constants.
We identify the mesonic superfields with the Higgs superfields of the SUSY SM
 as well as extra iso-spin doublet and singlet chiral superfields in an extended Higgs sector,
 and indentify the effective superpotential with that of the extended Higgs sector.
In this way, we introduce, with a solid UV completion, extra superfields that couple with the Higgs superfields 
 through large coupling constants.
A feature of the model is that the scale at which the large coupling constants blow up
 is identified with the confinement scale of the SUSY gauge theory.

We investigate the phenomenology of the SUSY extended Higgs sector 
 that has been obtained as the low-energy effective theory of the SUSY SU(2) gauge theory.
With a benchmark mass spectrum,
 we calculate the strength of EWPT and look for parameter regions
 where strongly first order EWPT occurs.
We further calculate the triple coupling constant of the SM-like Higgs boson
 and the decay branching ratio into di-photon,
 whose deviations from the SM values give collider signatures of the model.
\\

\section{Model}
We consider a SUSY extended Higgs sector 
 that emerges as a low-energy effective theory of
 a new SUSY SU(2)$_H$ gauge theory with six doublet chiral superfields,
 which are also charged under SM gauge groups SU(2)$_L \times$ U(1)$_Y$.
A $Z_2$ parity is assigned to the doublets to forbid large flavor changing neutral currents
 in the resultant extended Higgs sector.
One can say that the SUSY SU(2)$_H$ gauge theory is the UV picture of the model.
The field content of the SUSY SU(2)$_H$ gauge theory is summarized in Table 1.
\begin{table}
\begin{center}
\begin{tabular}{|c|c|c|c|} \hline
Field                 & $SU(2)_{L}$ & $U(1)_{Y}$ & $Z_{2}$ \\ \hline
$\left(
\begin{array}{c}
T_{1}  \\
T_{2}
\end{array}
\right)$              & 2           & 0          & +       \\ \hline
$T_{3}$               & 1           & +1/2       & +       \\ \hline
$T_{4}$               & 1           & $-$1/2       & +       \\ \hline
$T_{5}$               & 1           & +1/2       & $-$       \\ \hline
$T_{6}$               & 1           & $-$1/2       & $-$       \\ \hline
\end{tabular}
\end{center}
\caption{
SM charge and $Z_2$ parity assignments on the $SU(2)_H$ doublets, $T_i$.
}
\end{table}

The SU(2)$_H$ gauge theory becomes strongly-coupled at an IR scale, 
 which we call `confinement scale' $\Lambda_H$,
 and below $\Lambda_H$, the low-energy effective theory is described in terms of
 mesonic chiral superfields.
In our model, the mesonic superfields 
 are identified with the Higgs doublets of the SUSY SM as well as
 extra chiral superfields in the extended Higgs sector.
One can say that the extended Higgs sector is the IR picture of the model.
The Higgs sector contains
 two SU(2)$_L$ doublet, two charged singlet and five neutral singlet chiral superfields,
 in addition to the two Higgs doublets of the SUSY SM.
The field content is summarized in Table 2.
\begin{table}
\begin{center}
\begin{tabular}{|c|c|c|c|} \hline
Field                 & $SU(2)_{L}$ & $U(1)_{Y}$ & $Z_{2}$ \\ \hline
$H_{u}$               & 2           & +1/2       & $+$      \\ \hline
$H_{d}$               & 2           & $-$1/2       & $+$       \\ \hline
$\Phi_{u}$            & 2           & +1/2       & $-$       \\ \hline
$\Phi_{d}$            & 2           & $-$1/2       & $-$       \\ \hline
$\Omega^+$               & 1           & +1       & $-$      \\ \hline
$\Omega^-$               & 1           & $-$1       & $-$       \\ \hline
$\zeta$, $\eta$          & 1           & 0       & $-$       \\ \hline
$N$, $N_{\Phi}$, $N_{\Omega}$ & 1           & 0       & $+$       \\ \hline
\end{tabular}
\end{center}
\caption{
The chiral superfields in the extended Higgs sector.
}
\end{table}

The scalar components of the superfields $N, N_{\Phi}, N_{\Omega}$ gain vacuum expectation values,
 which give effective $\mu$-terms for the other superfields.
The physical components of $N_{\Phi}, N_{\Omega}$
 do not contribute to the one-loop effective potential for the SUSY SM Higgs scalars,
 and hence can be neglected in phenomenological studies.
The superpotential of the phenomenologically relevant part of the Higgs sector 
 is then given by
\begin{eqnarray}
W_{Higgs} &=& -\mu H_uH_d-\mu_{\Phi}\Phi_u\Phi_d-\mu_{\Omega}(\Omega^+\Omega^- - \zeta\eta)
	\nonumber\\
	&+&\hat{\lambda} \ \left\{ \ n H_uH_d +
	H_d\Phi_u\zeta + H_u\Phi_d\eta - H_u\Phi_u\Omega^- - H_d\Phi_d\Omega^+ \ \right\} \label{superpot} \ ,
\end{eqnarray}
 where $n$ denotes the physical component of $N$.
$\hat{\lambda}$ denotes a running coupling constant for the superfields 
 in the extended Higgs sector.
Na$\ddot{\i}$ve Dimensional Analysis (NDA) \cite{nda} suggests that 
 the coupling constant $\hat{\lambda}$ becomes non-perturbative
 at a similar scale where the SUSY SU(2)$_H$ gauge theory in the UV picture
 becomes strongly-coupled, \textit{i.e.}, the confinement scale $\Lambda_H$.
Therefore, with a given renormalization group equation of $\hat{\lambda}$,
 the value of $\hat{\lambda}$ at the electroweak scale 
 is in one-to-one correspondence with the confinement scale $\Lambda_H$.

The soft SUSY breaking terms are introduced as follows:
\begin{eqnarray}
{\cal L}_{soft} &=& -m_{H_u}^2 H_u^{\dagger} H_u \ - \ m_{H_d}^2 H_d^{\dagger} H_d
        \ - \ m_{\Phi_u}^2 \Phi_u^{\dagger} \Phi_u \ - \ m_{\Phi_d}^2 \Phi_d^{\dagger} \Phi_d \nonumber
\\
        &-& m_{\Omega^+}^2 \Omega^{+ \, \dagger} \Omega^+ \ - \ m_{\Omega^-}^2 \Omega^{- \, \dagger} \Omega^-
        \ - \ m_{\zeta}^2 \zeta^{\dagger} \zeta \ - \ m_{\eta}^2 \eta^{\dagger} \eta \nonumber
\\
&-& B\mu H_u H_d \ - \ B\mu_{\Phi} \Phi_u \Phi_d \ - \ B\mu_{\Omega} ( \Omega^+ \Omega^- - \zeta \eta) \nonumber
\\
&-& A_{\zeta} H_d\Phi_u\zeta \ - \ A_{\eta} H_u\Phi_d\eta 
\ - \ A_{\Omega^-} H_u\Phi_u\Omega^- \ - \ A_{\Omega^+} H_d\Phi_d\Omega^+ \ .
\end{eqnarray}
Similar to the minimal SUSY SM, electroweak symmetry breaking occurs
 with the help of soft SUSY breaking terms.

The Yukawa couplings, including the top quark Yukawa coupling,
 are introduced in the same way as in the minimal fat Higgs model \cite{fat higgs}.
\\

\section{Electroweak Phase Transition in the Model}
Using finite-temperature effective potential and the methods introduced in
 \cite{Funakubo:2009eg, Parwani:1991gq},
 we make a numerical analysis on the order of EWPT,
 $v_C / T_C$, where $T_C$ and $v_C$ respectively denote the critical temperature
 and the value of the order parameter at that temperature.

The benchmark mass spectrum is as follows.
For the SUSY SM sector, 
\begin{eqnarray}
& & \tan \beta \ = \ 15 \ , \ \ \
m_{H^{\pm}} \ = \ 350 \ {\rm GeV} \ , \ \ \ 
\mu \ = \ 200 \ {\rm GeV} \ , \nonumber \\
& & \tilde{M}_{\tilde{t}} \ = \ \tilde{M}_{\tilde{b}} \ = \ 2000 \ {\rm GeV} \label{mssm sample} \ ,
\end{eqnarray}
 where $\tilde{M}_{\tilde{t}}$ and $\tilde{M}_{\tilde{b}}$ respectively denote
 the soft SUSY breaking masses for the SUSY tops and bottoms.
For the $Z_2$-odd sector,
\begin{eqnarray}
& & \mu_{\Phi} \ = \ \mu_{\Omega} \ = \ 550 \ {\rm GeV} \ , \nonumber
\\
& & \bar{m}_{\Phi_{d}} \ = \ \bar{m}_{\Omega^+} \ = \ \bar{m}_{\zeta} \ = \ 1500 \ {\rm GeV} \ , 
\ \ \ \bar{m}_{\eta} \ = \ 2000 \ {\rm GeV} \ , \nonumber 
\\
& & ({\rm A terms, \ B terms}) \ = \ 0 \label{z2odd sample} \ ,
\end{eqnarray}
 where $\bar{m}_{*}$ denotes the square root of the sum of the $\mu$-term squared and 
 the soft SUSY breaking mass squared for the field $*$.
The following two quantities are the free parameters in this analysis:
\begin{eqnarray}
\lambda \ (\ \equiv \ \hat{\lambda}(M_Z) \ ) \ , \ \ \ \ \  m_{0} \ (\ \equiv \ \bar{m}_{\Phi_{u}} \ = \ \bar{m}_{\Omega^-} \ ) \label{free} \ .
\end{eqnarray}
We tune the value of the stop mixing term
 to realize $m_h=126$ GeV.

\begin{figure}[htbp]
  \begin{center}
   \includegraphics[width=100mm]{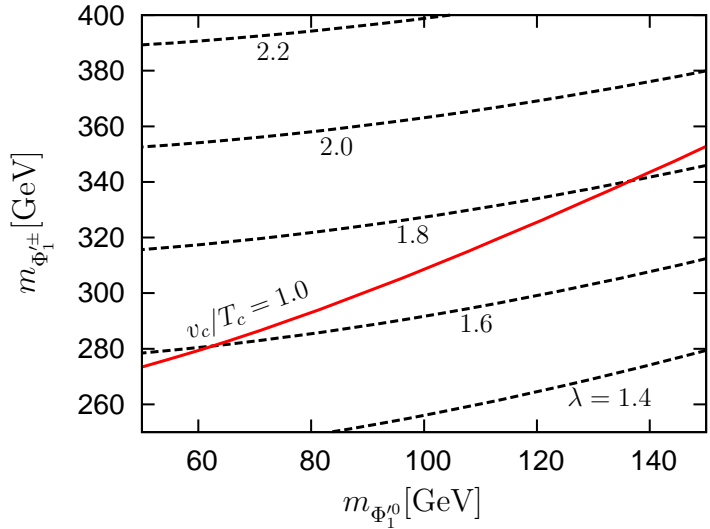}
  \end{center}
  \caption{Contour plot for the coupling constant $\lambda$ (black dashed lines) 
   with a line corresponding to the strength of EWPT $v_C/T_C=1$
   (red solid line),
   on the plane of 
   the mass of the lightest $Z_2$-odd \textit{charged} particle $m_{\Phi^{\prime \, \pm}_1}$
   and the mass of the lightest $Z_2$-odd \textit{neutral} particle $m_{\Phi^{\prime \, 0}_1}$.
  The parameters are fixed according to eqs.~(\ref{mssm sample}) and (\ref{z2odd sample}).
  }
  \label{vctc}
\end{figure}

In Figure~\ref{vctc}, we show the contour plot for the coupling constant 
 $\lambda \ (\ \equiv \ \hat{\lambda}(M_Z) \ )$.
The strength of EWPT, $v_C/T_C=1$, is also displayed.
We find that
 strongly first order phase transition, $v_C/T_C \gtrsim 1$, takes places
 with our benchmark mass spectrum
 for $\lambda \gtrsim 1.6$ when $m_{\Phi^{\prime \, 0}_1} \simeq 60$ GeV
 (for $\lambda \gtrsim 1.8$ when $m_{\Phi^{\prime \, 0}_1} \simeq 130$ GeV).
Note that, in our model, 
 $\lambda \simeq 1.6$ corresponds to $\Lambda_H \simeq 15$ TeV
 and $\lambda \simeq 1.8$ does to $\Lambda_H \simeq 5$ TeV.
\\

\section{Collider Signatures of the Model}
With the benchmark mass spectrum eqs.~(\ref{mssm sample}, \ref{z2odd sample}, \ref{free}),
 we make a numerical analysis on the decay branching ratio of the Higgs boson into diphoton
 and the triple Higgs boson coupling,
 and study the correlation between these quantities and the strength of EWPT.

\begin{figure}[htbp]
  \begin{center}
   \includegraphics[width=100mm]{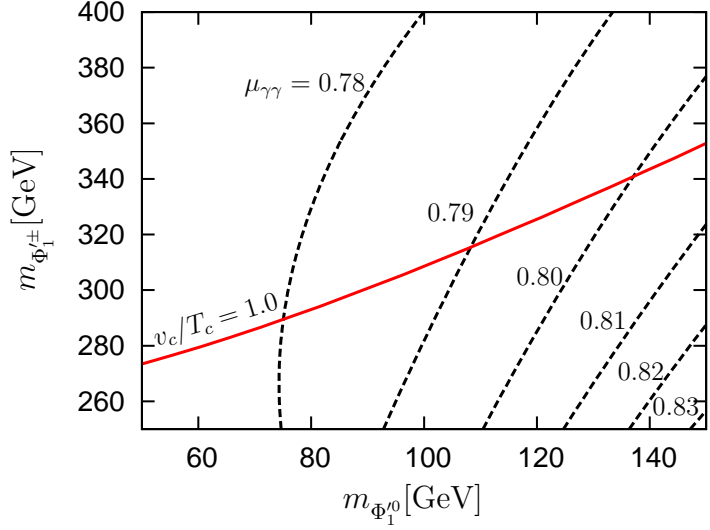}
  \end{center}
  \caption{Contour plot for the ratio of $Br(h \rightarrow \gamma \gamma)$ 
   over the SM value, $\mu_{\gamma \gamma}$ (black dashed lines),
   with a line corresponding to the strength of EWPT $v_C/T_C=1$
   (red solid line), 
   on the plane of
   the mass of the lightest $Z_2$-odd \textit{charged} particle $m_{\Phi^{\prime \, \pm}_1}$
   and the mass of the lightest $Z_2$-odd \textit{neutral} particle $m_{\Phi^{\prime \, 0}_1}$.
  The parameters are fixed according to eqs.~(\ref{mssm sample}) and (\ref{z2odd sample}).
  }
  \label{vctchgg}
\end{figure}

In Figure~\ref{vctchgg}, we combine the contour plot for 
 the ratio of the Higgs-to-diphoton branching ratio over its SM value, 
 $\mu_{\gamma \gamma}$,
 with a line indicating the strength of EWPT, $v_C/T_C=1$.
We find that the Higgs-to-diphoton branching ratio decreases by more than 20 \% 
 with our benchmark mass spectrum when the strongly first order EWPT with $v_C/T_C \gtrsim1$
 is realized.

\begin{figure}[tbp]
  \begin{center}
   \includegraphics[width=100mm]{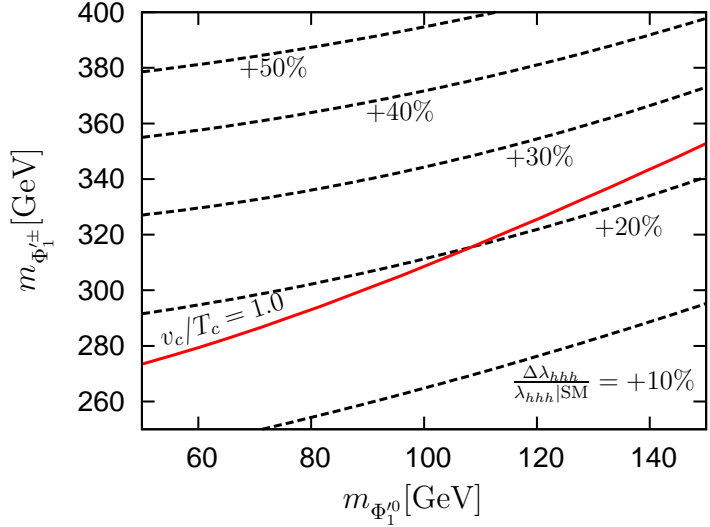}
  \end{center}
  \caption{Contour plot for the deviation of the triple Higgs boson coupling from the SM value, 
   $\Delta \lambda_{hhh}/\lambda_{hhh}\vert_{SM}$
   (black dashed lines),
   with a line corresponding to the strength of EWPT $v_C/T_C=1$
   (red solid line), on the plane of
   the mass of the lightest $Z_2$-odd \textit{charged} particle $m_{\Phi^{\prime \, \pm}_1}$
   and the mass of the lightest $Z_2$-odd \textit{neutral} particle $m_{\Phi^{\prime \, 0}_1}$.
  The parameters are fixed according to eqs.~(\ref{mssm sample}) and (\ref{z2odd sample}).
  }
  \label{vctchhh}
\end{figure}

In Figure~\ref{vctchhh}, we combine the contour plot for 
 the deviation of the triple Higgs boson coupling from the SM value,
 $\Delta \lambda_{hhh}/\lambda_{hhh}\vert_{SM}$,
 with a line indicating the strength of EWPT, $v_C/T_C=1$.
We discover that,
 when the strongly first order EWPT with $v_C/T_C \gtrsim 1$ occurs
 with our benchmark spectrum,
 the triple Higgs boson coupling increases by more than about 20 \% 
 for 150 GeV $> m_{\Phi^{\prime \, 0}_1} >$ 50 GeV.

To summarize, we confirm that sufficiently strongly first order EWPT for
 successful EWBG
 can be realized with our benchmark mass spectrum.
In order to have $v_C/T_C \gtrsim 1$,
 we need $\lambda > 1.6$
 provided the lightest $Z_2$-odd neutral scalar is heavier than 50 GeV.
This corresponds to the confinement scale $\Lambda_H$ lower than about 15 TeV.
In the parameter regions where strongly first order EWPT occurs,
 the Higgs-to-diphoton branching ratio, $Br(h \rightarrow \gamma \gamma)$,
 and the triple Higgs boson coupling, $\lambda_{hhh}$, significantly deviate
 from the SM values.
These are principally due to loop corrections involving light $Z_2$-odd scalars,
 which are also responsible for strongly first order EWPT.
With the benchmark mass spectrum,
 $Br(h \rightarrow \gamma \gamma)$ decreases by about 20\%
 and $\lambda_{hhh}$ increases by more than about 20\%,
 both of which may be observed at the future International Linear Collider \cite{peskin, fujii}.
\\

\section{Conclusions}
We have discussed the correlation among the strength of EWPT,
 the Higgs-to-diphoton branching ratio and the triple Higgs boson coupling
 in the extended Higgs sector with large coupling constants and the 126 GeV Higgs boson,
 which emerges as a low-energy effective theory of
 the SUSY SU(2)$_H$ gauge theory with confinement.
In our benchmark mass spectrum,
 the condition of quick sphaleron decoupling for EWBG, $v_C/T_C \gtrsim 1$,
 determines the scale of the Landau pole 
 to be below about 15 TeV, 
 which corresponds to the confinement scale of the SU(2)$_H$ gauge theory.
We have found that the Higgs-to-diphoton branching ratio deviates negatively from the SM prediction
 by about 20\%
 and the triple Higgs boson coupling deviates positively by more than about 20\%.
Such deviations can be observed at future collider experiments.
\\

\section*{Acknowledgment}

TY thanks Shinya Kanemura (University of Toyama), Eibun Senaha (KIAS) 
 and Tetsuo Shindou (Kogakuin University)
 for their collaboration on writing the papers~\cite{ours1, ours2}.
\\

\end{document}